\documentclass[12pt]{article}
\PassOptionsToPackage{table}{xcolor}
\usepackage{graphicx}
\usepackage{amsmath,amsthm,amssymb,mathrsfs,mathtools,bm}
\usepackage{tabularx}
\usepackage{booktabs}
\usepackage[OT1]{fontenc}
\usepackage{etoc}
\usepackage{float}
\usepackage{titlesec}
\usepackage{multirow,array}
\usepackage{diagbox}
\usepackage[top=1.25in, bottom=1.25in,left=1.25in,right=1.25in]{geometry}
\usepackage{fancyhdr}

\titleformat{\section}
		{\scshape \large \bfseries\center}
         {\thesection}
        {0.5em}
        {}
        []
\titleformat{\subsection}
        {\normalfont\bfseries}
        {\thesubsection}
        {0.5em}
        {}
        []

\allowdisplaybreaks

\def\subsectiontitle{}
\def\subsubsectiontitle{}
\usepackage{comment}
\usepackage[backgroundcolor=orange!50!white]{todonotes}
\usepackage{tikz}
\usetikzlibrary{automata, positioning}

\usepackage{kpfonts}
\usepackage{inconsolata}

\makeatletter
\@ifpackageloaded{unicode-math}{\let\bm\symbf}{}
\makeatother

\usepackage{makecell}
\usepackage{cellspace}
\usepackage[shortlabels]{enumitem}

\usepackage[colorlinks,breaklinks=true]{hyperref}
\usepackage{breakcites}

\usepackage[table]{xcolor}
\AtBeginDocument{%
	\hypersetup{%
    linkcolor={red!50!black},%
    citecolor={blue!50!black},%
    urlcolor={blue!80!black}%
}%
}%

\usepackage[nameinlink,noabbrev,sort,capitalise]{cleveref}
\makeatletter
\def\ps@pprintTitle{%
 \let\@oddhead\@empty
 \let\@evenhead\@empty
 \def\@oddfoot{\emph{Very preliminary version}\hfill\emph{This draft: \today}}%
 \let\@evenfoot\@oddfoot}
\makeatother
\usepackage{etoolbox}
\patchcmd{\pprintMaketitle}
  {\fi\hrule}
  {\fi\ifvoid\extrainfobox\else\unvbox\extrainfobox\par\vskip10pt\fi\hrule}
  {}{}

\newsavebox\extrainfobox

\newtheorem{prop}{Proposition}
\crefname{prop}{Proposition}{Propositions}

\newtheorem{thm}{Theorem}
\crefname{thm}{Theorem}{Theorems}

\crefname{cor}{Corollary}{Corollaries}

\newtheorem{lem}{Lemma}
\crefname{lem}{Lemma}{Lemmas}

\newtheorem{ass}{Assumption}
\crefname{ass}{Assumption}{Assumptions}

\crefname{axiom}{Axiom}{Axioms}

\newtheorem{defi}{Definition}
\crefname{defi}{Definition}{Definitions}

\theoremstyle{remark}

\crefname{remark}{Remark}{Remarks}

\theoremstyle{claim}

\crefname{claim}{Claim}{Claims}

\theoremstyle{definition}
\newtheorem{eg}{Example}
\crefname{eg}{Example}{Examples}

\crefname{problem}{Problem}{Problems}

\usepackage{indentfirst}

\allowdisplaybreaks

\let\oldfootnote\footnote
\renewcommand\footnote[1]{\oldfootnote{\hspace{.4mm}#1}}

\renewcommand{\footnotesize}{\scriptsize}

\makeatletter
\renewenvironment{proof}[1][\proofname] {\par\pushQED{\qed}\normalfont\topsep6\p@\@plus6\p@\relax\trivlist\item[\hskip\labelsep\bfseries#1\@addpunct{.}]\ignorespaces}{\popQED\endtrivlist\@endpefalse}
\makeatother

\let\oldFootnote\footnote
\newcommand\nextToken\relax

\renewcommand\footnote[1]{%
    \oldFootnote{#1}\futurelet\nextToken\isFootnote}

\newcommand\isFootnote{%
    \ifx\footnote\nextToken\textsuperscript{,}\fi}

\usepackage[american]{babel}
\usepackage[authoryear,round]{natbib}
\def\d{\mathrm{d}}
\newcommand{\rp}[1]{\langle #1 \rangle}
\def\E{\mathbb{E}}
\def\R{\mathbb{R}}
\def\H{\mathrm{Hess}}

\DeclareMathOperator*{\argmax}{argmax}
\DeclareMathOperator*{\argmin}{argmin}
\usepackage{todonotes}

\titlespacing\section{0pt}{6pt}{4pt}
\titlespacing\subsection{0pt}{4pt}{2pt}
\titlespacing\subsubsection{0pt}{2pt}{2pt}
 \titlespacing*{\paragraph}{0pt}{1.25ex plus 1ex minus .2ex}{0.5em}

\newenvironment{assumption}{\begin{ass}}{\end{ass}}
\newenvironment{definition}{\begin{defi}}{\end{defi}}
\newenvironment{proposition}{\begin{prop}}{\end{prop}}
\newenvironment{theorem}{\begin{thm}}{\end{thm}}
\newenvironment{lemma}{\begin{lem}}{\end{lem}}

\newenvironment{APPENDICES}{\appendix}{}
\newcommand{\FIGURE}[3]{\centering #1\caption{#2}#3}
\newcommand{\Halmos}{}

\makeatletter
\renewenvironment{proof}[1]
  {\par\pushQED{\qed}\normalfont\topsep6\p@\@plus6\p@\relax
   \trivlist\item[\hskip\labelsep\bfseries#1\@addpunct{.}]\ignorespaces}
  {\popQED\endtrivlist\@endpefalse}
\makeatother
\begin{document}

\title{Pricing Access to Dynamic Information Services}
\date{}
\author{\makebox[.25\linewidth]{Weijie Zhong\thanks{Graduate School of Business, Stanford University; email: \protect\texttt{weijie.zhong@stanford.edu}}}}
\maketitle

\begin{abstract}
A provider sells a \emph{dynamic information service}---a real-time, capacity-constrained process that resolves a customer's uncertainty---to customers who differ privately in urgency. I characterize the revenue-optimal mechanism: deploy a \emph{single}, undistorted information process---the one a customer with unlimited access would most prefer---and screen entirely through a one-dimensional menu of service caps. The provider leaves the product itself undistorted, unlike a Mussa--Rosen monopolist; all screening is absorbed into the cap. The mechanism rationalizes a recurring contractual form---a flat fee for capped access to a common process---spanning AI service tiers, expert-network consultations, and analyst-inquiry retainers. I identify the economic force behind the result, a convexity-preservation property of urgency screening, and give conditions under which the per-unit price declines in the cap.
\end{abstract}

\noindent\textbf{Keywords:} screening; nonlinear pricing; dynamic information design; AI subscription pricing; usage caps; service tiers; preference alignment.

\newpage
\section{Introduction}\label{sec:intro}

Consider two subscribers of the same AI assistant. One, an analyst building next quarter's forecasting model, is rarely in a hurry: an answer that arrives overnight serves as well as an answer now. The other, a reporter racing a filing deadline, asks questions of equal stakes whose answers are worthless by evening. The analyst buys the top plan---a large usage window at a higher fee and a lower price per unit of use---and lets long explorations run as far as they need; the reporter buys the entry plan and moves on whenever the window cuts an exploration short. The model serving both is identical; the plans differ only in how much of it each may consume before the window resets. How should a provider of a real-time information service price access when customers differ in how urgently they need answers? And why do the menus observed in practice vary caps but not the model itself?

The pattern is common. Consumers of AI services pay for capped access windows on a shared model;\footnote{A rolling five-hour message quota at Anthropic Pro or Max \citep{anthropic-2026a}; a three-hour rolling quota of flagship-model messages at ChatGPT Plus \citep{openai-2026a}; a monthly allotment of generation credits at Google AI Ultra \citep{google-2026}.} developers buying the same providers' APIs choose among service tiers---OpenAI's Priority, Standard, Flex, and Batch; Anthropic's Priority and Batch---that differ only in how the latency clock runs on the \emph{identical} underlying model \citep{openai-2026b,anthropic-2026b}. The pattern extends well beyond AI: expert networks (GLG, AlphaSights) sell prepaid packages of consultation-hour credits; industry-analyst subscriptions (Gartner, Forrester) bundle flat-fee access with capped inquiry time; fractional general-counsel retainers post tiered schedules of monthly legal-counsel hours. In each case the provider commits a \emph{capacity} to produce knowledge in real time and the customer buys a right of access to it: a fixed model behind a usage cap, expert matching behind a block of consultation hours, analyst expertise behind capped inquiry time, legal advice behind monthly counsel-hours.

A justification for these contracts still eludes nonlinear-pricing theory. When customers differ, the classical prescription \citep{mussa-rosen-1978,maskin-riley-1984} is to \emph{distort the product}: offer degraded variants so that less valuable types sort onto worse versions. These providers do the opposite---they preserve a single common service and screen through \emph{access rights}: caps, deadlines, priority classes. The answer developed here turns on two features special to information-service markets: the information process itself can be designed nearly freely, and the salient private dimension is the customer's \emph{urgency}.

I study this market. A monopolist provider sells a \emph{dynamic information service}: she designs a posterior belief process $\rp{\mu_t}$ that gradually resolves a customer's uncertainty about an unknown state, subject to a per-time \emph{throughput} constraint that bounds how fast information can be produced. The design is fully flexible: the provider may commit to any belief process satisfying the constraint and could, in principle, give each type a different one. Flexibility is realistic for these services, and it is the right theoretical benchmark: restricting attention to a familiar family prejudges the answer, as the two designs below illustrate. A customer's urgency is her private type: type $r$ earns $e^{-r\tau}$ when the process resolves the state at the random time $\tau$, where the discount rate $r$ measures how heavily she discounts a delayed answer. The provider chooses a menu of processes and transfers to maximize revenue subject to incentive compatibility and individual rationality. I state the problem with a single customer of privately known type; equivalently, the provider faces a unit mass of customers whose urgencies are distributed according to $G$.

Two designs that look natural for such a service illustrate the crux. The provider could sell \emph{committed delivery}---a guaranteed answer at a promised time---but under a hard throughput constraint early certainty is infeasible, and the best committed-delivery menu collapses to a single take-it-or-leave-it offer of the earliest delivery time. She could instead sell \emph{diffusive exploration}---a menu of learning speeds, the canonical Gaussian model of gradual learning---which also collapses to one offer. The optimum is neither: one breakthrough process offered to everyone behind a menu of caps out-earns both, as the running example of \cref{eg:1} quantifies. The gap is the value of designing the information process rather than picking it from a familiar family.

\paragraph{Main result.} The optimum collapses this rich design space to a single dimension. The provider deploys \emph{one} information process, the \emph{greedy exploration process}: at every instant it concentrates the entire throughput budget on the states currently closest to the customer's belief, so that a breakthrough---a signal that jumps the belief to an answer and ends the service---arrives at the highest feasible rate; absent a breakthrough, the belief drifts and the exploration widens to new candidate states. Greedy exploration is the process a customer with no cap would design for herself. Screening runs solely through a menu of \emph{time caps}: type $r$ pays $P(r)$ for the right to consume the common process up to a type-specific time $T(r)$. The process is identical for every type, and all screening is absorbed into the one-dimensional cap, whose per-unit price falls as the cap grows. The managerial principle: \textbf{to screen urgency, ration access to the service rather than degrade it.}

To understand why one process serves every customer, note first that the cap self-selects: a customer who cannot wait gains little from service arriving after her deadline, so she takes a small cap at a high per-unit price, while a patient customer takes a large cap at a low one. What has to be checked is that the provider never gains by also bending the process itself toward some type. \cref{prop:1} delivers this: the greedy exploration process maximizes $\E[\rho(\tau)]$ for \emph{every} nonnegative, decreasing, convex time preference $\rho$, and each customer's time preference, adjusted for the information rents her report commands, is one such function wherever serving her is profitable. One process is therefore optimal for all types at once, and only the cap varies; \cref{sec:analysis} develops the underlying convexity-preservation argument. The greedy process coincides with what the machine-learning literature calls a \emph{preference-aligned} output\footnote{\label{fn:rlhf}The terminology tracks the Reinforcement Learning from Human Feedback paradigm \citep{christiano-et-al-2017,ouyang-et-al-2022,bai-et-al-2022,stiennon-et-al-2020}, under which commercial language models are post-trained to produce outputs matching human-revealed preferences. The result here rationalizes one facet of that design target, non-distortion of the process; it is silent on the broader notions of helpfulness and harmlessness.}---the process a customer would choose for herself---so the profit motive aligns with this facet of alignment.

Three implications follow for providers of such services. First, profit maximization alone rationalizes deploying the customer-preferred process: degrading the model for cheaper tiers loses revenue rather than protecting it. Second, urgency is screened through caps alone, so separating impatient from patient customers requires no product line---one product and a menu of quotas suffice. Third, the per-unit price should fall as the cap grows, matching the volume discounts on observed menus.

\paragraph{Scope.} The model speaks to the pricing of capacity-constrained dynamic information services. It explains caps on a common service, priority and batch tiers, inquiry-hour bundles, prepaid expert-service credits, and the lower per-unit prices that latency-patient customers receive. It abstracts from forces that operate on other margins---multi-model menus, tiers driven by usage volume or task value, open-ended generation without a well-defined state, and prompt engineering---which I take up in \cref{sec:application}.

\paragraph{Related literature.} The paper joins classical screening and nonlinear pricing \citep{mussa-rosen-1978,maskin-riley-1984,myerson-1981,armstrong-1996,rochet-chone-1998,armstrong-vickers-2001,laffont-martimort-2002} to the literature on selling information and experiments \citep{bergemann-bonatti-2015,bergemann-bonatti-2019,bergemann-bonatti-smolin-2018,horner-skrzypacz-2016,yang-2022}. The classical literature screens on an exogenous quality or quantity; the screening instrument here---the stopping-time distribution of a dynamically evolving belief process---is new, and relative to Mussa--Rosen the result is a dimensionality collapse onto a scalar cap. The selling-information literature characterizes optimal menus of posteriors or experiments in static frames; I sell a real-time information \emph{process}, with \citet{horner-skrzypacz-2016}'s gradual revelation the closest antecedent, the difference being that the customer's private information is time preference rather than the state.

The technical home is dynamic information acquisition and optimal stopping, where a marketing literature characterizes the \emph{demand} side of dynamic information---a decision maker optimally consuming a given information technology. \citet{ke-villasboas-2019} study gradual learning across alternatives before choice, with consideration sets forming endogenously; \citet{ke-shen-villasboas-2016} characterize search across multiple products under Brownian information flows, finding that a seller would facilitate information for low-valuation consumers and obfuscate it for high-valuation ones; \citet{chaimanowong-ke-villasboas-2026} allocate a flow of learning capacity across projects when news arrives as Poisson breakthroughs. Economics counterparts include \citet{fudenberg-strack-strzalecki-2018} on speed--accuracy tradeoffs, \citet{morris-strack-2019} on sequential sampling costs, \citet{zhong-2019-jmp} on optimal dynamic Poisson experimentation, \citet{liang-mu-2020} on the learning traps that complementarities among information sources create, \citet{liang-mu-syrgkanis-2022} on dynamically aggregating diverse sources and the optimality of myopic attention strategies, and \citet{che-mierendorff-2019} and \citet{che-kim-mierendorff-2023} on dynamic attention and disengagement. This paper solves the corresponding \emph{supply} side: the provider designs the information process itself and prices access to it---the question \citet[p.~411]{ke-villasboas-2019} pose in closing, asking how a monopolist should choose ``information to provide, and prices to charge''---and the optimal design turns out to be the breakthrough learning that the demand-side models take as given, screened by caps.

The throughput constraint has an information-theoretic pedigree. With Shannon entropy it caps the service's mutual-information rate \citep{shannon-1948}; with general $H$ it is a flow version of the posterior-separable information costs of rational inattention \citep{sims-2003,caplin-dean-2015,matejka-mckay-2015,caplin-dean-leahy-2022,hebert-woodford-2022}, in the uniformly posterior-separable class that \citet{bloedel2020cost} microfound as the cost of optimally acquired information. In that literature the constraint prices information inside a decision maker's mind, as a cognitive cost; here the same object is a production technology---the provider's capacity---and, as \cref{sec:framework} notes, the reason urgency is screenable at all.

A separate literature designs information to steer receivers rather than to sell it. Optimal persuasion \citep{kamenica-gentzkow-2011} can often be implemented as a single public experiment \citep{kolotilin-et-al-2017}, including with privately informed or heterogeneous receivers \citep{alonso-camara-2016,bardhi-guo-2018}---a collapse my single-process result echoes. In marketing, \citet{yao-2024} studies dynamic persuasion of a consumer who can also search on his own, and \citet{ke-lin-lu-2026} study a platform that deliberately coarsens a public signal to limit how far consumers search, monetizing the information indirectly through commissions and advertising. Like the platform, my provider profits by limiting information consumption; unlike it, she prices access directly, and the restriction is a menu of caps that customers self-select into.

The closest mechanism-design antecedent is \citet{bergemann-bonatti-smolin-2026}, who screen AI customers on task value through an aggregation lemma that collapses the type to a scalar, obtaining a rich quantity-distortion menu of committed-spend and two-part-tariff contracts; I screen on urgency in a dynamic setting and obtain a single common process behind a one-dimensional cap. The two mechanisms differ in private type, instrument, and timing, and together they cover the two principal screening axes in commercial AI pricing: committed spend and quotas on one side, rolling-window caps and latency tiers on the other. Finally, the greedy process coincides with what the machine-learning literature calls \emph{preference-aligned} output (Footnote~\ref{fn:rlhf}); \cref{thm:1} supplies an economic rationale for one facet of it, non-distortion of the process, distinct from \emph{value alignment} in the AI-safety sense \citep{amodei-et-al-2016,hadfield-menell-hadfield-2018}.

\cref{sec:framework} introduces the framework. \cref{sec:analysis} characterizes the optimal mechanism. \cref{sec:ext} develops extensions to heterogeneous valuations and endogenous reasoning quality. \cref{sec:application} maps the model to AI and other knowledge-service pricing and marks the mechanism's boundaries. \cref{sec:conclusion} concludes. Proofs are in the appendix.

\section{Framework}\label{sec:framework}

\subsection{Setup}\label{sec:setup}

A customer (\emph{he}) would like to learn an unknown state $\theta \in \Theta = \{1, \ldots, n\}$ with common prior $\mu_0 \in \Delta(\Theta)$. If the state is learned at time $t \in \R_+$, the customer obtains utility $e^{-rt}$, where the discount rate $r \in [\underline{r}, \overline{r}]$, $0 < \underline{r} < \overline{r}$, is the customer's privately known \emph{urgency} parameter. The distribution of $r$ has CDF $G$ and PDF $g$, known to the provider. I impose the standard monotone reverse hazard rate condition: $r \mapsto g(r)/G(r)$ is weakly decreasing on $(\underline r, \overline r]$. I frame the model with a single customer of private type; equivalently, the provider faces a unit mass of customers whose urgencies are distributed by $G$, and the direct mechanism below is written for a representative customer.

A provider (\emph{she}) designs a \emph{belief process} $\rp{\mu_t}_{t \geq 0}$ on $\Delta(\Theta)$, a c\`adl\`ag martingale by Bayes' rule. The design is fully flexible: any martingale satisfying the throughput constraint below is available. Flexibility has two justifications. Practically, the information policies of the services I model are themselves designed objects---an AI model's behavior is shaped end to end by training, fine-tuning, and decoding choices. Theoretically, imposing an ad hoc parametric family can mislead: in \cref{eg:1}, two natural families each collapse to take-it-or-leave-it pricing and forfeit most of the attainable revenue, so full flexibility is the right benchmark. The process ends when the state is learned:
\begin{align*}
\tau(\mu) = \inf\{t \geq 0 : \mu_t \in \{e_\theta\}_{\theta=1}^n\},
\end{align*}
where $e_\theta = (0, \ldots, 0, 1, 0, \ldots, 0)$ is the degenerate belief at state $\theta$.

\subsection{Information Throughput}\label{sec:info-throughput}

The provider's capacity limits how fast she can produce information: compute bounds what an AI model generates per unit time, and a professional's working hours bound what an expert, analyst, or lawyer delivers in a day. The model captures this as an \emph{information-throughput} constraint on the belief process: for all $t, s \geq 0$,
\begin{align}
\E\!\left[H(\mu_{t+s}) - H(\mu_t) \mid \mathcal{F}_t\right] \leq \chi s, \tag{Info.}\label{eq:info}
\end{align}
where $H : \Delta(\Theta) \to \R$ is a strictly convex (generalized entropy) function. The expected increase in $H$ measures informational content generated; the constraint bounds it by $\chi > 0$, the throughput rate. \cref{sec:mapping} discusses the technological origin of $\chi$ in AI applications.

The constraint is what makes urgency matter: without it the provider could resolve the state at $t=0$, so every type would obtain payoff $1$ and urgency would be irrelevant. Screening rents then arise from customers' private information about the urgency that the constraint renders payoff-relevant.

Let $\mathcal{M}$ denote the set of c\`adl\`ag martingale processes satisfying \eqref{eq:info}. I assume $H$ is $C^{(2)}$ smooth and \emph{irreducible}.\footnote{At every interior $\mu$, the graph on $\Theta$ that places an edge between $\theta \neq \theta'$ whenever $\H H(\mu)_{\theta\theta'} \neq 0$ is connected. Irreducibility is the standard matrix notion; it says the entropy couples all states.} Without loss, extend $H$ from $\Delta(\Theta)$ to the nonnegative orthant homogeneously of degree one, with $H(0):=0$; the extension is convex there, being positively homogeneous and subadditive (subadditivity is convexity of $H$ on the simplex applied to the normalized sum), and satisfies Euler's identities $\nabla H(\mu) \cdot \mu = H(\mu)$ and $\H H(\mu)\, \mu = \mathbf{0}$. Let $D$ be the Bregman divergence associated with $H$:
\begin{align*}
D(\mu' \mid \mu) := H(\mu') - H(\mu) - \nabla H(\mu) \cdot (\mu' - \mu).
\end{align*}

\begin{assumption}\label{ass:1}
The divergence $D$ satisfies:
\begin{enumerate}[(i)]
\item $\sup_{\mu \in \Delta(\Theta)} \min_{\theta \in \Theta} D(e_\theta \mid \mu) < \infty$;
\item there exists $\epsilon > 0$ such that for all $\mu \in \Delta(\Theta)$ with $\mu(\theta) \leq \epsilon$, $D(e_\theta \mid \mu) > \min_{\theta' \neq \theta} D(e_{\theta'} \mid \mu)$;
\item for all interior $\mu$ and all $\theta \neq \theta'$, $(e_\theta - \mu)^\top \nabla_\mu D(e_{\theta'} \mid \mu) \geq 0$.
\end{enumerate}
\end{assumption}

\cref{ass:1} imposes three regularity properties on $D$. Condition (i) bounds the divergence to the ``closest state,'' preventing the process from stalling at interior beliefs from which no terminal state is reachable at positive rate. Condition (ii) rules out a near-zero-probability state being Bregman-closest: if the belief on $\theta$ falls below $\epsilon$, some $\theta' \neq \theta$ must be closer than $\theta$. Condition (iii) is a \emph{monotone-repulsion} property---moving the belief toward $\theta$ never decreases the divergence to any other state $\theta'$---equivalently, $(e_\theta-\mu)^\top \H H(\mu)\,(e_{\theta'}-\mu) \leq 0$ for $\theta \neq \theta'$, so $\H H(\mu)$ has nonpositive off-diagonal entries. Together with the irreducibility of $\H H$, condition (iii) makes the Hessian an irreducible singular $M$-matrix, which (\cref{sec:proof-prop1}) gives the throughput-allocation system below a unique \emph{positive} solution.

The Kullback--Leibler divergence (from Shannon entropy), with Hessian $\H H(\mu) = \mathrm{diag}(1/\mu) - \bm{1}\bm{1}^\top$, satisfies \cref{ass:1} in every dimension (\cref{sec:appendix-assumption}) and is the leading example. The quadratic (Mahalanobis) divergence satisfies \cref{ass:1} in the binary case, where condition (iii) is automatic; it enters the paper only through the binary examples.

\subsection{Mechanism}\label{sec:mechanism}

A (direct) mechanism specifies, for each reported type $r'$, (i) a belief process $\rp{\mu^{r'}_t} \in \mathcal{M}$ and (ii) a transfer $P(r')$. Let $U(r' \mid r) = \E[e^{-r \tau(\mu^{r'})}]$ denote the utility of type $r$ reporting $r'$. The provider solves
\begin{align}
\max_{\rp{\mu^r} \in \mathcal{M},\, P} \int_{\underline r}^{\overline r} P(r)\, g(r)\, \d r \tag{P}\label{prob:1}
\end{align}
subject to incentive compatibility, $U(r \mid r) - P(r) \geq U(r' \mid r) - P(r')$ for all $r, r'$, and individual rationality, $U(r \mid r) - P(r) \geq 0$ for all $r$.

Throughout, the provider is assumed to commit to the announced mechanism. In the AI application, commitment corresponds to a published service tier that cannot be unilaterally renegotiated within the contract window; relaxing commitment is a natural direction for future work.

The timing is as follows. First, the provider posts the menu of plans, one per report. Second, the customer privately observes his urgency $r$ and selects a plan or walks away. Third, the plan's information process runs until it resolves the state or is exhausted, and the posted transfer is paid.

\subsection{Example}\label{eg:1}

I illustrate with a binary setting, which I carry through the paper as a running example. Let $\theta \in \{0, 1\}$ with prior $\mu_0 = 1/2$, and $r \sim U[1, 2]$. Take quadratic-variation throughput: $\E[(\mu_{t+s} - \mu_t)^2 \mid \mathcal{F}_t] \leq s/8$ (here \cref{ass:1} holds because $n=2$). The example does double duty: it fixes ideas, and it shows how two designs that look natural for an information service mislead. Both \emph{committed delivery} (a guaranteed delivery time) and \emph{diffusive exploration} (the canonical Gaussian learning model) are feasible and intuitive, yet each forces take-it-or-leave-it pricing and leaves revenue on the table relative to the flexible optimum of \cref{sec:analysis}, a qualitatively different process.

\paragraph{Committed delivery.} Suppose the provider commits to resolve the answer deterministically at a delivery time $t$, selling the contract $(t, P)$. Feasibility under \eqref{eq:info} constrains the \emph{resolution time}, not any particular sample path: producing the information needed to resolve a symmetric binary state---posterior variance $1/4$---at rate $\chi = 1/8$ takes time at least $2$, so a deterministic resolution time $t$ is feasible if and only if $t \geq 2$.\footnote{The implementing process produces information at the full rate throughout $[0,t]$: its accumulated entropy rises to the maximum exactly at $t$. Since a $[0,1]$-valued martingale with mean $1/2$ attains the maximal variance $1/4$ only at the degenerate law on $\{0,1\}$, such a process resolves the state at exactly $t$. (Such a deterministic resolution time is realized by a gradual full-throughput process. See \citealp{chen2025information}.)} A type-$r$ customer's utility from contract $(t, P)$ is $U(t \mid r) = e^{-rt} - P$, decreasing and log-linear in $t$ with slope $-r$. Selling a menu $\{(t(r), P(r))\}_{r \in [1,2]}$ is therefore a one-dimensional Mussa--Rosen problem in the single quantity $t$, with single-crossing in $(r, t)$ and feasibility bound $t \geq 2$. The optimum is a take-it-or-leave-it offer at $t = 2$ serving customers with $r \leq 1.5$ and excluding the rest; revenue $\approx 0.025$.

\paragraph{Diffusive exploration.} Now suppose the belief evolves as Brownian motion $\d\mu^\sigma_t = \sigma\, \d B_t$ until first hitting $\{0, 1\}$, indexed by volatility $\sigma$---the canonical Gaussian model of gradual learning. The process satisfies \eqref{eq:info} iff $\sigma^2 \leq 1/8$. Solving the associated HJB for the value of learning the state at the first-hitting time gives the type-$r$ utility $U(\sigma \mid r) = \mathrm{sech}(\sqrt{r/2}/\sigma)$, increasing in $\sigma$ with single-crossing in $(r, \sigma)$. Selling a menu $\{(\sigma(r), P(r))\}$ is again a one-dimensional Mussa--Rosen problem, this time in $\sigma$. The optimum is a single take-it-or-leave-it offer at the maximum feasible volatility $\sigma^* = 1/(2\sqrt{2})$, yielding revenue $\mathrm{sech}(2\sqrt{2}) \approx 0.12$---five times committed delivery.

Both designs force take-it-or-leave-it pricing. The question is whether the provider gains by designing over the full set $\mathcal{M}$. \cref{thm:1} answers in the affirmative: the optimal design is a qualitatively different belief process---neither committed delivery nor diffusive exploration---priced with a non-trivial menu, yielding revenue $\approx 0.2$, eight times committed delivery and twice diffusive exploration. \cref{sec:eg-revisited} gives closed forms. The lesson generalizes: the revenue-optimal information product is a process that no familiar parametric family delivers, so guessing within one understates what flexible design achieves---and, as \cref{sec:analysis} shows, the optimal process takes the same shape for every type.

\section{Analysis}\label{sec:analysis}

I solve \eqref{prob:1} in three steps: reduce to a pointwise problem via the envelope theorem; characterize the pointwise optimum via a greedy exploration process; verify global IC and IR.

\subsection{Pointwise Reduction}\label{sec:pointwise}

Fix a menu $(\rp{\mu^r}, P)$. The envelope theorem gives
\begin{align*}
\frac{\d}{\d r}\big(U(r \mid r) - P(r)\big) = \E\big[-e^{-r\tau(\mu^r)}\, \tau(\mu^r)\big],
\end{align*}
and hence
\begin{align*}
P(r) = \E\big[e^{-r\tau(\mu^r)}\big] - \int_r^{\overline r} \E\big[-e^{-z\tau(\mu^z)}\, \tau(\mu^z)\big]\, \d z - \big[U(\overline r \mid \overline r) - P(\overline r)\big].
\end{align*}

Setting $U(\overline r \mid \overline r) - P(\overline r) = 0$, revenue becomes
\begin{align}
\int_{\underline r}^{\overline r} \E\!\left[e^{-r\tau(\mu^r)}\left(1 - \tau(\mu^r)\frac{G(r)}{g(r)}\right)\right] g(r)\, \d r. \tag{Relaxed}\label{eq:relaxed}
\end{align}

This is separable in $r$, so \eqref{eq:relaxed} decomposes into pointwise maximization of the \emph{virtual time preference} $\rho_r(t) := e^{-rt}(1 - t\, G(r)/g(r))$ over belief processes in $\mathcal{M}$.

\subsection{Greedy Exploration under Convex Time Preference}\label{sec:greedy}

I solve the pointwise problem by constructing and verifying a conjectured solution. Before the formal definition, here is the process in three steps.
\begin{enumerate}[topsep=2pt,itemsep=1pt,label=(\arabic*)]
\item \emph{Aim at the closest states.} At each instant the process identifies the terminal states closest to the current belief in Bregman divergence and directs \emph{all} throughput at them.
\item \emph{Wait for a breakthrough.} A breakthrough about each targeted state arrives at a Poisson rate, jumping the belief to certainty and ending the service. Absent a breakthrough, the belief drifts and new states gradually join the closest set.
\item \emph{Once all states are in play, hold steady.} When every state is equidistant, the belief stops drifting and the process becomes a stationary Poisson race in which each state breaks through at a rate proportional to its probability, scaled to spend the full budget.
\end{enumerate}
The name \emph{greedy} reflects step~(1): allocating all throughput to the closest states maximizes the instantaneous arrival rate of a resolving signal.

\begin{definition}[Greedy exploration process]\label{defi:greedy}
The greedy exploration process $\rp{\mu^*_t}$ is defined recursively. Initialize $k = 1$, $\widehat{\mu}^1 = \mu_0$, $\widehat t^1 = 0$, and $\Theta^1 = \argmin_{\theta \in \Theta} D(e_\theta \mid \widehat \mu^1)$.
\begin{itemize}
\item While $\Theta^k \subsetneq \Theta$, for $t \geq \widehat t^k$, define Poisson rates $(\beta_t(\theta))_{\theta \in \Theta^k}$ and path $\widehat \mu_t$ by
\begin{align}
& D(e_\theta \mid \widehat \mu_t) = D(e_{\theta'} \mid \widehat \mu_t), \quad \forall\, \theta, \theta' \in \Theta^k; \tag{iso-D}\label{eq:isoD}\\
& \sum_{\theta \in \Theta^k} \beta_t(\theta)\, D(e_\theta \mid \widehat \mu_t) = \chi; \tag{info}\label{eq:info-rate}\\
& \frac{\d \widehat \mu_t}{\d t} = -\sum_{\theta \in \Theta^k} \beta_t(\theta)\, (e_\theta - \widehat \mu_t); \tag{drift}\label{eq:drift}
\end{align}
with initial condition $\widehat \mu_{\widehat t^k} = \widehat \mu^k$. Let $\widehat t^{k+1}$ be the earliest time at which a new state enters $\argmin_\theta D(e_\theta \mid \widehat \mu_t)$; update $\widehat \mu^{k+1} = \widehat \mu_{\widehat t^{k+1}}$, $\Theta^{k+1} = \argmin_\theta D(e_\theta \mid \widehat \mu^{k+1})$; increment $k$.
\item Once $\Theta^k = \Theta$, set $\widehat \mu_t \equiv \widehat \mu^k$ for $t \geq \widehat t^k$ and, writing $\zeta := D(e_\theta \mid \widehat \mu^k)$ (common across $\theta$ by \eqref{eq:isoD}), set the stationary rates $\beta_t(\theta) \equiv \dfrac{\chi}{\zeta}\,\widehat \mu^k(\theta)$.
\end{itemize}
For $t \in [\widehat t^k, \widehat t^{k+1})$, the belief evolves as the compensated Poisson process
\begin{align*}
\d \mu^*_t = \sum_{\theta \in \Theta^k}\left(\d Q^\theta_t(\beta_t(\theta)) - \beta_t(\theta)\, \d t\right)(e_\theta - \mu^*_t),
\end{align*}
where $Q^\theta_t(x)$ are independent Poisson counters with arrival rate $x$.
\end{definition}

Two remarks on the construction. First, in the final phase the rates $(\chi/\zeta)\,\widehat\mu^k(\theta)$ keep the belief stationary---they are proportional to $\widehat\mu^k$, so the drift in \eqref{eq:drift} vanishes---while exhausting the budget, $\sum_{\theta} \beta_t(\theta)\, D(e_\theta \mid \widehat\mu^k) = (\chi/\zeta)\sum_\theta \widehat\mu^k(\theta)\,\zeta = \chi$; this is the $n$-state analog of the symmetric Poisson counter in the binary example below. Second, greedy exploration is \emph{myopically} optimal---it maximizes the instantaneous arrival rate of a resolving signal---but it is not optimal for a customer with concave time preference. It is revenue-maximizing precisely because the virtual time preferences generated by urgency screening inherit convexity, which is the content of \cref{prop:1}. Breakthrough learning of this conclusive-Poisson form is the information technology that demand-side search models take as given---most recently \citet{chaimanowong-ke-villasboas-2026}, who justify the specification by appeal to an optimizing learner; here it emerges as the revenue-optimal \emph{design}. \cref{fig:simplex} traces the process on a three-state simplex.

\begin{figure}[htbp]
\FIGURE
{
\begin{tikzpicture}[>=stealth,scale=1.35]
\draw[thick] (0,0) -- (4,0) -- (2,3.464) -- cycle;
\node[below left]  at (0,0)     {\footnotesize $e_1$};
\node[below right] at (4,0)     {\footnotesize $e_2$};
\node[above]       at (2,3.464) {\footnotesize $e_3$};
\fill[blue] (1.000,0.346) circle (1.5pt);
\node[below] at (1.02,0.30) {\footnotesize $\mu_0$};
\draw[very thick,blue,->] (1.000,0.346) -- (2.000,0.693);
\draw[dashed,blue,->] (1.000,0.346) to[bend left=18] (0.10,0.07);
\node[blue,align=center] at (0.62,1.05) {\footnotesize phase 1:\\[-3pt]\footnotesize explore state $1$};
\fill[red!80!black] (2.000,0.693) circle (1.5pt);
\draw[very thick,red!80!black,->] (2.000,0.693) -- (2.000,1.155);
\draw[dashed,red!80!black,->] (2.000,0.693) to[bend left=20] (0.14,0.03);
\draw[dashed,red!80!black,->] (2.000,0.693) to[bend right=20] (3.86,0.03);
\node[red!80!black,align=center] at (3.30,1.02) {\footnotesize phase 2:\\[-3pt]\footnotesize explore states $\{1,2\}$};
\fill[green!45!black] (2.000,1.155) circle (1.8pt);
\draw[dashed,green!45!black,->] (2.000,1.155) to[bend left=14] (0.16,0.10);
\draw[dashed,green!45!black,->] (2.000,1.155) to[bend right=14] (3.84,0.10);
\draw[dashed,green!45!black,->] (2.000,1.155) -- (2.000,3.30);
\node[green!45!black,align=center] at (2.02,2.06) {\footnotesize phase 3: explore all states;\\[-3pt]\footnotesize belief holds, rates $\propto\widehat\mu(\theta)$};
\end{tikzpicture}}
{The greedy exploration process on the belief simplex, $n=3$, KL divergence, prior $\mu_0=(0.7,0.2,0.1)$. The process aims all throughput at the currently closest states---under KL, the most likely ones. Solid segments trace the no-breakthrough drift; the dashed arrows at each phase's starting belief point to the states being explored, to which a breakthrough would jump. In phase 1 (blue) the process explores state $1$, and the belief drifts away from $e_1$ until state $2$ is equally close; in phase 2 (red) it explores states $\{1,2\}$ until state $3$ joins; in phase 3 (green) all states are explored, the belief holds at the barycenter. \label{fig:simplex}}
{}
\end{figure}

\begin{proposition}[Greedy exploration is optimal for every convex time preference]\label{prop:1}
For any nonnegative, decreasing, continuous, convex function $\rho : \R_+ \to \R_+$,\footnote{On paths that never resolve the state, $\rho(\tau)$ is read as $\lim_{t\to\infty}\rho(t)$; see \cref{sec:appendix-duality}.}
\begin{align}
\rp{\mu^*_t} \in \argmax_{\rp{\mu_t} \in \mathcal{M}} \E[\rho(\tau(\mu))]. \label{eq:prop1}
\end{align}
\end{proposition}

\cref{prop:1} is the technical pivot of my main result. The envelope-derived virtual time preference $\rho_r$ is \emph{convex on its positive region} (and can be freely truncated where negative), so the same greedy process maximizes expected virtual surplus \emph{for every type simultaneously}; only the truncation time varies.

\paragraph{Proof strategy.} The proof (\cref{sec:appendix-duality}) is an optimality certificate built from two elementary directions of the exploration--stopping duality of \citet{sannikov-zhong-2024}, both reproduced for completeness. \emph{(i) An outer bound:} every feasible process induces a stopping law satisfying an information-accounting inequality, so the optimum over the larger set of laws meeting that inequality bounds the optimum over feasible processes (\cref{lem:necessity}). \emph{(ii) A dual certificate:} an explicitly constructed multiplier pair bounds the value of that relaxation by weak duality (\cref{lem:duality}). The greedy process is feasible, induces a law meeting the inequality with equality, and attains the dual bound (\cref{lem:attain}); hence it is optimal. \footnote{I never invoke the converse (sufficiency) direction or strong duality from \citet{sannikov-zhong-2024}, because the primal process \emph{and} the dual certificate are constructed by hand---which is also why the strong-convexity hypothesis of \citet{sannikov-zhong-2024} plays no role here.} The construction extends the binary-state result of \citet{chen2025information} to general finite state spaces under \cref{ass:1}.

\subsection{Optimal Mechanism}\label{sec:optmech}

Let $T(r) := g(r)/G(r)$. The virtual time preference $\rho_r(t) = e^{-rt}(1 - t/T(r))$ is strictly convex and positive for $t < T(r)$ and negative for $t > T(r)$. Its truncation $\rho_r^+(t) := \max\{\rho_r(t), 0\} = \rho_r(\min\{t, T(r)\})$ is nonnegative, continuous, weakly decreasing, and weakly convex (the kink at $T(r)$ is upward in slope since $\rho_r'(T(r)^-) < 0 = \rho_r'(T(r)^+)$); \cref{prop:1}, extended by mollification (\cref{lem:moll}), thus applies, so $\rp{\mu^*_t}$ maximizes $\E[\rho_r^+(\tau(\mu))]$. Define the $T(r)$-truncation $\mu^{*T(r)}_t := \mu^*_{\min\{t, T(r)\}}$. Then $\rp{\mu^{*T(r)}_t}$ solves the pointwise problem.

\begin{definition}[Service-quota menu]\label{defi:token-menu}
The service-quota menu consists of:
\begin{itemize}
\item A single greedy exploration process $\rp{\mu^*_t}$, inducing the stopping-time distribution $f^*$;
\item A menu $\{(\chi T(r), P(r))\}_{r \in [\underline r, \overline r]}$, where the cap is $\chi T(r) = \chi\, g(r)/G(r)$ and the transfer is
\begin{align*}
P(r) = \int_0^{T(r)} e^{-rt} f^*(t)\, \d t - \int_r^{\overline r} \int_0^{T(z)} e^{-zt}\, t\, f^*(t)\, \d t\, \d z.
\end{align*}
\end{itemize}
\end{definition}

I use \emph{time cap} for the time-units quantity $T(r)$ and \emph{service quota} for the information-units quantity $\chi T(r)$. The service quota is the contractible object observed in practice---a token allowance, a spending budget, or a query quota, depending on the application. The marginal price per additional unit of cap is
\begin{align*}
\frac{P^{*\prime}(r)}{T'(r)} = e^{-r T(r)} f^*(T(r)).
\end{align*}
In words, the marginal unit of quota is priced at the customer's discounted probability that this unit is the one that resolves his question: $f^*(T(r))$ is the chance that resolution arrives exactly at the cap, and $e^{-rT(r)}$ is what that resolution is worth to him. The marginal price is \emph{decreasing} in the cap $\chi T(r)$ whenever (a) the greedy stopping density $f^*$ is weakly decreasing and (b) $r\,T(r)$ is weakly decreasing in $r$. Condition (a) holds automatically: along the greedy path the no-breakthrough drift moves the belief \emph{away} from the closest states, so $\zeta(t) = \min_\theta D(e_\theta \mid \mu^*_t)$ is non-decreasing and the breakthrough hazard $\chi/\zeta(t)$ non-increasing (immediate on single-active-state segments from positive definiteness of $\H H$; verified for the KL and quadratic divergences in \cref{sec:appendix-assumption}). Condition (b) is a mild restriction on $G$ beyond the monotone reverse hazard rate.\footnote{Parametrizing by the cap $c = T(r)$ and the type $r(c)$, the marginal price is decreasing in the cap if and only if $\d \log f^*(c)/\d c \leq \d\bigl(r(c)\,c\bigr)/\d c$. Condition (b) does not follow from the monotone reverse hazard rate: a slowly decaying $T(r)$ can make the per-unit price rise locally even when $f^*$ is decreasing.} Both hold in the binary example of \cref{sec:eg-revisited}, where $f^*$ is exponential and $r\,T(r) = r/(r-1)$ is decreasing.

\begin{theorem}[One process, a menu of caps]\label{thm:1}
The service-quota menu is an optimal mechanism for \eqref{prob:1}.
\end{theorem}

\begin{proof}{Proof.}
\cref{prop:1} implies the menu attains the value of the relaxed program \eqref{eq:relaxed}. For global IC, verify supermodularity of $U(r' \mid r)$ in $(r, r')$:
\begin{align*}
\frac{\partial^2}{\partial r\, \partial r'} \int_0^{T(r')} e^{-rt} f^*(t)\, \d t = -T'(r')\, T(r')\, e^{-r T(r')}\, f^*(T(r')) \geq 0,
\end{align*}
using $T'(r) \leq 0$ under the monotone reverse hazard rate assumption. Payments are determined by local IC. Under the monotone reverse hazard rate, $T(r)$ is decreasing, so the type with the lowest surplus from participation is $\overline r$ (both the discount factor $e^{-rt}$ and the cap $T(r)$ are decreasing in $r$); setting $P(\overline r) = U(\overline r \mid \overline r)$ binds IR with equality there, and the envelope formula delivers $P(r)$ for all $r < \overline r$.
\Halmos\end{proof}

\cref{thm:1} has three interpretations worth flagging.

\paragraph{(i) Profit motive produces a user-optimal process.} The single belief process deployed by the revenue-optimal mechanism maximizes the expected utility of a customer with no cap; it is identical to the process a planner maximizing gross user surplus would design. The provider's only lever against the customer is the one-dimensional cap; the shape of the process is common to every type. This coincidence of revenue-maximizing and surplus-maximizing design is special: a Mussa--Rosen monopolist screening on quality \emph{does} distort the product itself. It arises here because convexity of the customer's time preference is preserved under the envelope-induced adjustment, collapsing the infinite-dimensional design problem to the one-dimensional cap. The machine-learning literature calls the surplus-maximizing process \emph{preference-aligned} (see Footnote~\ref{fn:rlhf}) and trains models to produce it. \cref{thm:1} supplies an economic rationale for \emph{one} aspect of this design target---non-distortion of the belief process itself---with no appeal to intrinsic commitments to user welfare. The result is silent on other aspects of preference alignment (helpfulness, honesty, refusal behavior) that require richer preference structures than the discount-rate model considered here.

\paragraph{(ii) Dimensionality collapse.} A priori the infinite-dimensional design space admits a menu of type-customized processes, as multidimensional screening would suggest. \cref{thm:1} shows the problem \emph{collapses}: the optimal mechanism uses none of the process-shape dimensions and loads all screening onto the scalar cap $T(r)$, distorted downward for every type except the most patient---exactly as the scalar quantity is in Mussa--Rosen \citep{mussa-rosen-1978,maskin-riley-1984}. The novelty is the collapse itself: the \emph{information-production technology} is left undistorted, and the cap---an access right the customer does experience---carries the screening. This parallels \citet{kolotilin-et-al-2017}, where optimal persuasion collapses to a single public experiment; the aggregation lemma of \citet{bergemann-bonatti-smolin-2026}, where multidimensional task value collapses to a scalar; and \citet{ke-lin-lu-2026}, where a platform's information-design problem over posterior vectors collapses to the choice of a scalar search length. The collapse is special to convex time preference: $\rho_r$ stays convex under the envelope-induced adjustment, so one process is optimal for every type's \emph{virtual} preference.

Two features deserve flagging. First, the comparative statics are \emph{inverted} relative to Mussa--Rosen. There, higher types have both higher total and higher marginal valuation; here, for a deterministic resolution time $t$, the most patient (low-$r$) type has the \emph{highest} total valuation $e^{-rt}$ but the \emph{lowest} marginal valuation for speed---so it is the urgent types whose consumption is truncated. Second, all types participate: $T(\overline r) = g(\overline r) > 0$ (since $G(\overline r) = 1$), so even the most urgent type receives a positive cap and binds individual rationality with equality rather than being excluded, in contrast to the restricted families of \cref{eg:1} where feasibility alone forces take-it-or-leave-it pricing; and the most patient type $\underline r$ faces no cap at all ($T(r) \to \infty$ as $G(r) \to 0$), the no-distortion-at-the-top analog.

\paragraph{(iii) Quota pricing rationalized.} The optimal menu---a common process screened by a one-dimensional cap, with per-unit price decreasing in the cap under the conditions above---matches the structure of pricing across a range of dynamic knowledge services: AI subscriptions and API service tiers sell access to a common model differentiated by caps on stopping time, which practice implements through usage windows and latency tiers, and the same template appears in expert-network consultation packages, analyst-inquiry retainers, and fractional general-counsel agreements. \cref{sec:application} develops the mapping and the forces behind the features the theorem leaves out---chiefly the multi-model menus offered alongside cap-screened tiers, which I attribute to forces orthogonal to urgency screening.

\subsection{Example Revisited}\label{sec:eg-revisited}

In the binary setting, $D(0 \mid 0.5) = D(1 \mid 0.5) = 1/4$. The greedy process is therefore a stationary Poisson counter: the belief stays at $0.5$ or jumps to $\{0, 1\}$ with total rate $\chi / (1/4) = 1/2$ (each state at rate $1/4$). The stopping distribution is exponential: $f^*(t) = \tfrac{1}{2} e^{-t/2}$. The cap is $\chi T(r) = \tfrac{1}{8(r-1)}$. Optimal revenue:
\begin{align*}
\pi^* = \int_1^2 \int_0^{1/(r-1)} \tfrac{1}{2} e^{-(r+1/2)t}(1 - t(r-1))\, \d t\, \d r \approx 0.2,
\end{align*}
eight times committed delivery and twice diffusive exploration.

\cref{fig:spine} summarizes the mechanism in this binary example: panel (a) shows the common greedy process and the type-dependent caps that truncate it; panel (b) shows the resulting menu---the cap and the per-unit price as functions of urgency---and how patient and urgent customers sort.
\begin{figure}[htbp]
\FIGURE
{
\begin{tikzpicture}[>=stealth]
\begin{scope}
  \draw[->] (0,0) -- (5.4,0) node[right] {\footnotesize time $t$};
  \draw[->] (0,0) -- (0,3.3) node[above] {\footnotesize belief $\mu_t$};
  \node[left] at (0,0)   {\scriptsize $0$};
  \node[left] at (0,1.5) {\scriptsize $\tfrac12$};
  \node[left] at (0,3)   {\scriptsize $1$};
  \draw[dotted] (0,3) -- (5,3);
  \draw[very thick] (0,1.5) -- (2.7,1.5);
  \draw[very thick,->] (2.7,1.5) -- (2.7,3);
  \draw[thick,dashed] (2.7,1.5) -- (2.7,0.05);
  \node[right] at (2.8,2.8) {\scriptsize Poisson breakthrough};
  \draw[dashed] (1.1,0) -- (1.1,1.5);
  \draw[dashed] (4.2,0) -- (4.2,1.5);
  \node[below] at (1.1,0) {\scriptsize $T(\overline r)$};
  \node[below] at (4.2,0) {\scriptsize $T(r')$};
  \node[align=center] at (1.1,1.85) {\scriptsize urgent};
  \node[align=center] at (4.2,1.85) {\scriptsize patient};
  \node at (2.5,-0.95) {\footnotesize (a) one common process, type-specific caps};
\end{scope}
\begin{scope}[xshift=8cm]
  \draw[->] (0,0) -- (3.6,0) node[right] {\footnotesize urgency $r$};
  \draw[->] (0,0) -- (0,3.3);
  \node[below] at (0.15,0) {\scriptsize $\underline r$};
  \node[below] at (3,0)    {\scriptsize $\overline r$};
  \draw[very thick] (0.2,2.9) to[out=-72,in=165] (3,0.55);
  \draw[very thick,dashed] (0.2,0.45) to[out=22,in=205] (3,2.35);
  \draw[very thick] (1.75,3.15) -- (2.15,3.15);
  \node[right] at (2.18,3.15) {\scriptsize  service quota};
  \draw[very thick,dashed] (1.75,2.82) -- (2.15,2.82);
  \node[right] at (2.18,2.82) {\scriptsize per-unit price};
  \node at (1.7,-0.95) {\footnotesize (b) the cap--price menu and sorting};
\end{scope}
\end{tikzpicture}}
{The mechanism in the binary running example ($\mu_0=\tfrac12$, $r\sim U[1,2]$, $\chi=\tfrac18$). \textbf{(a)} A common greedy process: the belief holds at $\tfrac12$ and resolves via a Poisson breakthrough (stopping density $f^*(t)=\tfrac12 e^{-t/2}$); the most impatient type's cap $T(\overline r)$ truncates early, a more patient type's cap $T(r')$ lies farther right, and the most patient type faces no cap at all ($T(\underline r)=\infty$). \textbf{(b)} The menu: service quota and per-unit price against urgency $r$; patient customers buy larger caps at lower per-unit prices. \label{fig:spine}}
{}
\end{figure}

\section{Extensions}\label{sec:ext}

\subsection{Heterogeneous Valuations}\label{sec:het-val}

Suppose customers also differ in valuation for the output: type-$r$ utility is $q(r)\, e^{-rt}$ with $q : [\underline r, \overline r] \to \R_+$, $C^{(1)}$. The virtual time preference becomes
\begin{align*}
q(r)\, e^{-rt}\left(1 - \big(t - q'(r)/q(r)\big)\frac{G(r)}{g(r)}\right),
\end{align*}
strictly positive and convex for $t < T(r) := g(r)/G(r) + q'(r)/q(r)$. The single-process-plus-cap structure survives whenever $T(r)$ is decreasing. The supermodularity check
\begin{align*}
\frac{\partial^2}{\partial r\, \partial r'}\, U(r \mid r') \;\propto\; T(r') - q'(r)/q(r)
\end{align*}
is positive whenever $q' \leq 0$ (value weakly decreasing in urgency), or more generally whenever $\max_r q'(r)/q(r) < T(\overline r)$ ($q$ not too steeply increasing). Under these conditions, value heterogeneity does not change the main result.

\subsection{Endogenous Reasoning Quality}\label{sec:endog-quality}

Suppose the customer earns $V(\mu_\tau)$ at the stopping time, where $V : \Delta(\Theta) \to \R_+$ is continuous and convex. A standard microfoundation: if the customer chooses an action $a \in A$ with state-dependent payoff $u(a, \theta)$, then $V(\mu) = \max_a \sum_\theta \mu(\theta)\, u(a, \theta)$ is convex in $\mu$ by the envelope theorem. The baseline corresponds to $V(\mu) = \mathbf{1}\{\mu \in \{e_\theta\}\}$; $V$ allows the customer to derive partial value from intermediate beliefs.

The provider now chooses both the process and the (endogenous) stopping time. The pointwise problem is
\begin{align*}
\max_{\rp{\mu_t},\, \tau} \E\!\left[V(\mu_\tau)\, e^{-r\tau}\left(1 - \tau\, \frac{G(r)}{g(r)}\right)\right].
\end{align*}

I do not solve this problem in general; I offer one parametric specification as an illustration. With $V(\mu) = |\mu - 0.5|$ and $H(\mu) = |\mu - 0.5|^\alpha$ in the binary setting, an explicit solution exists: the reasoning quality $\kappa(t) := |\mu_t - 0.5|$ decays according to a closed form involving the Kummer confluent hypergeometric function.\footnote{The closed form is $\kappa(t) = \big(\tfrac{\alpha \chi}{(\alpha-1)(\alpha+1)}\, (T(r) - t)\, e^{-\alpha r(T(r)-t)}\, {}_1F_1(\alpha+1, \alpha+2, \alpha r(T(r)-t))\big)^{1/\alpha}$. It is a direct application of the stopping-control framework of \citet{sannikov-zhong-2024} to the binary-state parametric specification: setting $\kappa(t) := |\mu_t - 0.5|$ and exploiting the symmetry of the problem, \eqref{eq:info} and the HJB optimality condition combine into a linear second-order ODE for $\kappa(t)$ on the interior stopping region, which reduces to Kummer's confluent hypergeometric equation once the boundary condition $\kappa(T(r)) = 0$ is imposed.} \cref{fig:2} plots the resulting stopping boundaries $0.5 \pm \kappa(t)$ for several values of $r$. Two qualitative features emerge. First, a \emph{quality--delay tradeoff}: more urgent types receive lower answer quality in exchange for higher stopping rates. As $r$ rises, the provider sacrifices output accuracy to deliver answers faster---in AI services, reasoning-optimized models that spend more inference compute under longer thinking budgets behave this way, and an expert giving a quick verbal read versus a considered opinion is the same tradeoff. Second, for all types other than the most patient $\underline r$, answer quality decays to zero as $t \to T(r)$---yielding \emph{soft caps} in which output quality deteriorates near the cap rather than the service terminating abruptly. A provider can implement this with one core service plus inexpensive type-specific adjustments near the cap---for an AI service, post-training fine-tunes that degrade output as consumption approaches the quota---so the single-product structure survives. I present this as a suggestive illustration of how the single-process logic bends under endogenous quality, not as a general result.

\begin{figure}[htbp]
\FIGURE
{\tikzset{every picture/.style={line width=0.75pt}} 

\begin{tikzpicture}[x=0.75pt,y=0.75pt,yscale=-1,xscale=1]

\draw    (60.11,240) -- (419.67,240) ;
\draw [shift={(421.67,240)}, rotate = 180] [color={rgb, 255:red, 0; green, 0; blue, 0 }  ][line width=0.75]    (10.93,-3.29) .. controls (6.95,-1.4) and (3.31,-0.3) .. (0,0) .. controls (3.31,0.3) and (6.95,1.4) .. (10.93,3.29)   ;
\draw    (127.07,235) -- (127.07,240) ;
\draw    (194.02,235) -- (194.02,240) ;
\draw    (260.98,235) -- (260.98,240) ;
\draw    (327.93,235) -- (327.93,240) ;
\draw    (394.88,235) -- (394.88,240) ;
\draw    (60.11,40) -- (60.11,140) ;
\draw    (60.11,90) .. controls (104.53,91.25) and (124.61,104.75) .. (127.07,140) ;
\draw [color={rgb, 255:red, 189; green, 16; blue, 224 }  ,draw opacity=1 ]   (60.34,84.25) .. controls (127.96,84.75) and (192.33,95) .. (194.02,140) ;
\draw [color={rgb, 255:red, 65; green, 117; blue, 5 }  ,draw opacity=1 ]   (60.34,80.75) .. controls (198.93,80.75) and (256.33,93) .. (260.98,140) ;
\draw [color={rgb, 255:red, 0; green, 122; blue, 255 }  ,draw opacity=1 ]   (59.67,76.25) .. controls (318.11,78.75) and (321.33,90) .. (327.93,140) ;
\draw [color={rgb, 255:red, 208; green, 2; blue, 27 }  ,draw opacity=1 ]   (59.67,73.25) .. controls (391.09,74.25) and (393.33,84) .. (394.88,140) ;
\draw  [dash pattern={on 0.84pt off 2.51pt}]  (59.89,69.5) -- (412.52,69.5) ;
\draw    (60.11,239.75) -- (60.11,140) ;
\draw    (60.11,189.88) .. controls (104.53,188.63) and (124.61,175.16) .. (127.07,140) ;
\draw [color={rgb, 255:red, 189; green, 16; blue, 224 }  ,draw opacity=1 ]   (60.34,195.61) .. controls (127.96,195.11) and (193.33,185) .. (194.02,140) ;
\draw [color={rgb, 255:red, 65; green, 117; blue, 5 }  ,draw opacity=1 ]   (60.34,199.1) .. controls (198.93,199.1) and (256.33,188) .. (260.98,140) ;
\draw [color={rgb, 255:red, 0; green, 122; blue, 255 }  ,draw opacity=1 ]   (59.67,203.59) .. controls (318.11,201.1) and (319.33,190) .. (327.93,140) ;
\draw [color={rgb, 255:red, 208; green, 2; blue, 27 }  ,draw opacity=1 ]   (59.67,206.58) .. controls (391.09,205.59) and (388.33,199) .. (394.88,140) ;
\draw  [dash pattern={on 0.84pt off 2.51pt}]  (59.89,210.32) -- (412.52,210.32) ;
\draw    (424.33,104.5) -- (438.67,104.5) ;
\draw [color={rgb, 255:red, 189; green, 16; blue, 224 }  ,draw opacity=1 ]   (424.33,119) -- (438.67,119) ;
\draw [color={rgb, 255:red, 65; green, 117; blue, 5 }  ,draw opacity=1 ]   (424.33,132.5) -- (438.67,132.5) ;
\draw [color={rgb, 255:red, 0; green, 122; blue, 255 }  ,draw opacity=1 ]   (424.83,146) -- (439.17,146) ;
\draw [color={rgb, 255:red, 208; green, 2; blue, 27 }  ,draw opacity=1 ]   (424.33,159) -- (438.67,159) ;
\draw  [dash pattern={on 0.84pt off 2.51pt}]  (424.33,172.5) -- (438.67,172.5) ;
\draw  [line width=0.75]  (417.83,94.5) -- (495.67,94.5) -- (495.67,183.75) -- (417.83,183.75) -- cycle ;

\draw (40.33,133.9) node [anchor=north west][inner sep=0.75pt]  [font=\footnotesize]  {$0.5$};
\draw (33.33,84.9) node [anchor=north west][inner sep=0.75pt]  [font=\footnotesize]  {$0.75$};
\draw (47.33,36.9) node [anchor=north west][inner sep=0.75pt]  [font=\footnotesize]  {$1$};
\draw (34.33,183.4) node [anchor=north west][inner sep=0.75pt]  [font=\footnotesize]  {$0.25$};
\draw (46.33,232.4) node [anchor=north west][inner sep=0.75pt]  [font=\footnotesize]  {$0$};
\draw (122.83,243.4) node [anchor=north west][inner sep=0.75pt]  [font=\footnotesize]  {$2$};
\draw (189.33,243.33) node [anchor=north west][inner sep=0.75pt]  [font=\footnotesize]  {$4$};
\draw (256.83,243.83) node [anchor=north west][inner sep=0.75pt]  [font=\footnotesize]  {$6$};
\draw (322.33,242.83) node [anchor=north west][inner sep=0.75pt]  [font=\footnotesize]  {$8$};
\draw (385.83,242.33) node [anchor=north west][inner sep=0.75pt]  [font=\footnotesize]  {$10$};
\draw (417.83,245.33) node [anchor=north west][inner sep=0.75pt]  [font=\footnotesize]  {$t$};
\draw (23.83,131.83) node [anchor=north west][inner sep=0.75pt]  [font=\footnotesize]  {$\mu $};
\draw (443.83,98.4) node [anchor=north west][inner sep=0.75pt]  [font=\footnotesize]  {$r=3/2$};
\draw (443.83,112.9) node [anchor=north west][inner sep=0.75pt]  [font=\footnotesize,color={rgb, 255:red, 189; green, 16; blue, 224 }  ,opacity=1 ]  {$r=5/4$};
\draw (443.83,126.4) node [anchor=north west][inner sep=0.75pt]  [font=\footnotesize,color={rgb, 255:red, 65; green, 117; blue, 5 }  ,opacity=1 ]  {$r=7/6$};
\draw (444.33,139.9) node [anchor=north west][inner sep=0.75pt]  [font=\footnotesize,color={rgb, 255:red, 0; green, 122; blue, 255 }  ,opacity=1 ]  {$r=9/8$};
\draw (443.83,152.9) node [anchor=north west][inner sep=0.75pt]  [font=\footnotesize,color={rgb, 255:red, 208; green, 2; blue, 27 }  ,opacity=1 ]  {$r=1.1$};
\draw (443.83,166.4) node [anchor=north west][inner sep=0.75pt]  [font=\footnotesize]  {$r=1$};

\end{tikzpicture}}
{Consumption-based reasoning quality. \label{fig:2}}
{}
\end{figure}

\section{Applications}\label{sec:application}

The mechanism is a template for pricing capacity-constrained dynamic information services. I develop three markets: AI services, expert networks, and industry-analyst subscriptions. For each, I first explain why the model's features---flexible information design, a capacity constraint, private urgency---fit the market, and then map the mechanism to the observed contracts. Every market shows the same signature: a common product across tiers, screening through caps alone, and per-unit prices that decline in the cap. I close with the margins the mechanism leaves to other forces.

\subsection{Mapping the Primitives}\label{sec:mapping}

The belief process $\rp{\mu_t}$ is the customer's resolution of uncertainty as the service proceeds; I read it as a reduced form covering both the production and the consumption of answers. The throughput rate $\chi$ is the speed at which the provider produces information. In an AI service, $\chi$ is pinned by GPU hardware \citep{vaswani2017attention,dao-et-al-2022,kwon-et-al-2023-vllm}; tokens are the accounting unit for the information rate, and per-information-unit cost is roughly uniform across the tiers of a model generation, which justifies the zero-marginal-cost normalization.\footnote{A larger model delivers fewer but more informative tokens per unit time, a smaller model the reverse. Pretraining cost is sunk before the service interval, as is prompting, which sets the task primitives $(\Theta,\mu_0)$; heterogeneous prompting ability is heterogeneity in $\mu_0$. The throughput rate is set by the technology; the customer does not choose it.} In the professional services below, $\chi$ is a specialist's or analyst's working time. The time cap $T(r)$ and the service quota $\chi T(r)$ are the contractible access right---a rolling message window, a block of consultation hours, a bundle of inquiry or advisory hours---and the transfer $P(r)$ is the subscription fee or retainer.

\subsection{Three Markets}\label{sec:markets}

\paragraph{AI services.} The model's features fit generative-AI services directly. \emph{Flexible design:} a model's information policy is a designed object---the flexibility argued in \cref{sec:framework}---and preference-aligned training (Footnote~\ref{fn:rlhf}) delivers the customer-optimal path. \emph{Capacity:} GPU throughput binds inference and pins $\chi$, as \cref{sec:mapping} records. \emph{Private urgency:} how quickly a customer needs answers---the analyst and the reporter of the opening example---is the customer's information alone.

The contracts match the menu. Consumer plans sell rolling usage windows on one model---the service quota $\chi T(r)$---at flat fees, with larger windows at higher fees and lower per-unit prices \citep{anthropic-2026a,openai-2026a,google-2026}. Developer tiers sell the same model at different effective delivery rates: Priority commits dedicated throughput at a premium, Standard is best-effort behind it, and Flex and Batch take leftover capacity at roughly half the synchronous price \citep{openai-2026b,openai-2026c,anthropic-2026b}; in the menu's terms, batch and flex occupy the large-cap, low-per-unit end chosen by latency-patient customers, and priority the small-cap, premium end. The tier distinction is a slowdown rather than a stop---Anthropic's documentation says ``processing may be slowed down based on current demand and your request volume'' \citep{anthropic-2026b}---and across every tier the same model is delivered, with the per-unit price declining in the allotment as \cref{sec:optmech} prescribes.

\paragraph{Expert networks.} GLG, AlphaSights, Third Bridge, and Guidepoint match clients with industry specialists for live consultations. \emph{Flexible design:} a consultation is steered in real time---the specialist pursues whatever line of questioning most advances the client's understanding, under no fixed protocol. \emph{Capacity:} the throughput is the specialist's attention; an hour of conversation transmits only so much. \emph{Private urgency:} whether the client is racing a deal deadline or building background knowledge is the client's information alone.

The commercial unit is a prepaid consultation-hour credit, roughly \$1{,}000--\$2{,}000 per hour, sold in annual packages with minimums near \$25{,}000--\$60{,}000 at premium networks \citep{inex-one-2026,woozle-2026}. The client draws down hours until the question is resolved or the credits run out. The specialist pool behind every package is the same, and larger packages carry lower per-hour rates---the common product and the declining per-unit schedule.

\paragraph{Industry-analyst subscriptions.} Gartner and Forrester sell access to analyst expertise. \emph{Flexible design:} an inquiry session goes wherever the client steers it, and the analyst tailors the follow-up rather than reading from a catalogue. \emph{Capacity:} analyst time is metered---each inquiry session is capped at 30 minutes, with ``extensive analysis or additional research'' excluded \citep{gartner-2026}. \emph{Private urgency:} a client in a live vendor negotiation needs the read this week; a client planning next year's architecture can wait a quarter.

Subscriptions bundle research access with capped inquiry time at \$25K--\$150K a year, with enterprise tiers higher \citep{vendr-2026}; advisory time sells separately in prepaid blocks, a direct $\{T,P\}$ menu with per-hour discounts at larger sizes \citep{vendr-2026,saleshive-2026}. The firm commits time rather than conclusions, and every tier draws on the same analyst bench.

\subsection{What the Mechanism Leaves to Other Margins}\label{sec:features-omitted}

The model isolates one screening margin. \emph{Urgency} here is the rate at which the value of a resolved answer decays, the discount rate $r$. It is a different margin from a priority premium for a fixed-value task---the queueing model of \citet{afeche-2013}, a complement---and from the volume or task value on which customers also differ. Because higher $r$ compresses the total value of the service, more urgent customers optimally receive smaller caps; the often-cited pattern that power users buy larger quotas is the volume/value margin that \citet{bergemann-bonatti-smolin-2026} screen on, while within the urgency margin the prediction is the batch/flex pattern of larger, cheaper-per-unit allotments for latency-patient customers. The demand side shows the same margin from the customer's chair: in \citet{ke-shen-villasboas-2016}'s search model, more impatient consumers optimally consider fewer products and search less before purchasing---the behavior my provider prices, and screens, from the supply side.

The clearest gap between model and data is the multi-model menu---GPT Nano/Mini/Pro, Claude Haiku/Sonnet/Opus, Gemini Flash/Pro---which a single-process result does not produce. These menus reflect across-tier forces orthogonal to urgency: pretraining-cost amortization and task specialization, delivered from a common GPU fleet but with distinct upstream investments. The pricing landscape thus separates into three layers: the within-tier \emph{urgency} layer this paper isolates; a \emph{value/volume} layer of committed-spend contracts and quotas, characterized by \citet{bergemann-bonatti-smolin-2026}; and a \emph{model-quality} layer of multi-model menus. The baseline also takes the task to have a well-defined state resolved in a single interaction, which leaves open-ended generation and multi-turn evolving-state conversations for future work.

\subsection{Implications for Pricing Managers}\label{sec:managerial}

The mechanism translates into four rules for a manager pricing a capacity-constrained knowledge service. \emph{Sell one product.} Deploy the process an uncapped customer would want---the preference-aligned model, the full-attention consultation---in every tier; a degraded variant for cheap tiers gives up revenue that caps protect. \emph{Screen urgency with access.} Caps, rolling windows, and delivery speed separate impatient from patient customers by themselves: the rolling windows, batch tiers, and prepaid credit packages of \cref{sec:markets} all screen access to an identical underlying service. \emph{Let the per-unit price fall in the cap.} The marginal unit is worth the customer's discounted chance that it resolves his question, which declines along the menu---the arithmetic behind batch discounts and bulk advisory-hour packages. \emph{Reserve product lines for other margins.} A multi-model lineup prices task value and volume, the margins of \cref{sec:features-omitted}; urgency alone never calls for one.

\section{Discussion and Conclusion}\label{sec:conclusion}

I characterize the revenue-optimal mechanism for a capacity-constrained dynamic information service sold to customers who differ privately in urgency: a \emph{single}, undistorted information process---the one an uncapped customer would choose---behind a one-dimensional menu of caps. Screening rations \emph{access} rather than degrading the \emph{service}, because the infinite-dimensional design problem collapses to scalar cap screening: convex time preference survives the envelope adjustment, so the customer-preferred, preference-aligned process is optimal for every type. The mechanism rationalizes the flat-fee-plus-cap contracts of AI service tiers, expert networks, and analyst subscriptions, and it is testable on posted menus: per-unit prices should decline in the cap, tiers should share the underlying product, and cap tightness should track latency sensitivity rather than volume.

The model's boundaries point to its extensions. The provider commits to posted plans, as published service tiers do; relaxing commitment would speak to negotiated enterprise contracts. The task is a single question with a well-defined answer; multi-turn interactions with evolving states, and joint screening of urgency with the value/volume margin of \citet{bergemann-bonatti-smolin-2026}, are the natural next steps. The throughput rate is a technological constant here; making it a choice variable would price the capacity investment itself.

\begin{APPENDICES}

\section{Feasibility and Weak Duality}\label{sec:appendix-duality}

The proof of \cref{prop:1} uses two results from \citet{sannikov-zhong-2024}: the \emph{necessity} of a capacity inequality, and \emph{weak} duality. Both are elementary directions, and I prove them here in the present finite-state setting so that the argument is self-contained. 

Fix a feasible process $\rp{\mu_t}\in\mathcal M$, adapted to a filtration satisfying the usual conditions, with $\mu_t=\Pr[\theta=\cdot\mid\mathcal F_t]$. Throughout, $\rho(\infty):=\lim_{t\to\infty}\rho(t)$---the limit exists, $\rho$ being decreasing and bounded below---and on paths that never resolve the state I use the convention $\E[\rho(\tau)]:=\int_0^\infty\rho\,\d F+\rho(\infty)\Pr[\tau=\infty]$. For each state $i$ set
\begin{align*}
F^i(t) := \Pr[\tau(\mu)\le t,\ \theta=i], \qquad F:=\textstyle\sum_i F^i,
\end{align*}
so that $F^i$ is the \emph{joint sub-distribution} of the stopping time on $\{\theta=i\}$---not a conditional CDF. Two identities follow from the martingale property. First, each coordinate $\mu_t(i)$ is a $[0,1]$-valued martingale, so the vertices are absorbing: on $\{\tau\le t\}$ the belief is frozen at a vertex, and a posterior of $e_i$ implies $\theta=i$ almost surely, whence $\E[\mu_t(i)\,\mathbf 1\{\tau\le t\}]=F^i(t)$. Second, $\E[\mu_t(i)]=\mu_0(i)$. Combining,
\begin{align*}
\mu_0(i)-F^i(t) = \Pr[\tau>t,\ \theta=i] = \E[\mu_t(i)\,\mathbf 1\{\tau>t\}],
\end{align*}
whether or not the process ever resolves the state; when it does not, the deficit $\mu_0(i)-F^i(\infty)$ is the mass on $\{\tau=\infty,\theta=i\}$. In particular $\sum_i(\mu_0(i)-F^i(t))e_i=(1-F(t))\,\bar\mu_t$, where $\bar\mu_t:=\E[\mu_t\mid\tau>t]$ is the survivor belief (defined when $F(t)<1$).

\begin{lemma}[Necessity / outer bound]\label{lem:necessity}
Every feasible process satisfies, for all $t\ge 0$,
\begin{align}
\sum_i F^i(t)\,H(e_i) + H\!\Bigl(\sum_i (\mu_0(i)-F^i(t))\,e_i\Bigr) - H(\mu_0) \le \chi\int_0^t (1-F(s))\,\d s. \tag{Cap}\label{eq:cap}
\end{align}
\end{lemma}
\begin{proof}{Proof.}
Let $A(t):=\E[H(\mu_t)]$; by absorption, $\mu_t=\mu_{t\wedge\tau}$ almost surely, so the stopped paths contribute the frozen value $H(e_\theta)$. Decompose the increment of $A$ on $\{\tau\le t\}$ and $\{\tau>t\}$: the first event contributes zero, and since $\{\tau>t\}\in\mathcal F_t$ while \eqref{eq:info} bounds the conditional increment at the deterministic pair $(t,t+s)$,
\begin{align*}
A(t+s)-A(t) = \E\bigl[\mathbf 1\{\tau>t\}\,\E[H(\mu_{t+s})-H(\mu_t)\mid\mathcal F_t]\bigr] \le \chi\,s\,\Pr[\tau>t].
\end{align*}
Hence $A$ is Lipschitz with $A'(t)\le \chi(1-F(t))$ almost everywhere, and $A(t)-H(\mu_0)\le \chi\int_0^t(1-F)$. Decomposing $A(t)$ itself on the same two events,
\begin{align*}
A(t)=\sum_i F^i(t)\,H(e_i) + \E[H(\mu_t)\,\mathbf 1\{\tau>t\}].
\end{align*}
If $F(t)<1$, Jensen (convexity of $H$) and degree-one homogeneity give $\E[H(\mu_t)\mathbf 1\{\tau>t\}] = (1-F(t))\,\E[H(\mu_t)\mid\tau>t] \ge (1-F(t))\,H(\bar\mu_t) = H\bigl(\sum_i(\mu_0(i)-F^i(t))e_i\bigr)$; if $F(t)=1$ both sides vanish, since $H(0)=0$. Combining the two displays gives \eqref{eq:cap}.
\Halmos\end{proof}

\cref{lem:necessity} yields a relaxation of the feasible set. Define
\begin{align*}
\mathcal R := \Bigl\{(F^\theta)_{\theta\in\Theta}:\ \text{each } F^\theta \text{ nondecreasing, c\`adl\`ag, } F^\theta(0^-)=0,\ F^\theta(\infty)\le\mu_0(\theta),\ \eqref{eq:cap}\ \forall t\ge 0\Bigr\},
\end{align*}
and for $(F^\theta)\in\mathcal R$ write $F:=\sum_\theta F^\theta$ and $m_\theta:=\mu_0(\theta)-F^\theta(\infty)$ for the mass never stopped. Extend each $F^\theta$ to a measure $\widetilde F^\theta$ on $(0,\infty]$ by placing the atom $m_\theta$ at $u=\infty$, and write $\int\rho\,\d\widetilde F:=\sum_\theta\int_{(0,\infty]}\rho(u)\,\d\widetilde F^\theta(u)$, evaluating $\rho$ at $u=\infty$ by $\rho(\infty)$. By \cref{lem:necessity} the stopping data of every feasible process lie in $\mathcal R$ with $\E[\rho(\tau)]=\int\rho\,\d\widetilde F$, so for every nonnegative decreasing $\rho$,
\begin{align}
\sup_{\rp{\mu_t}\in\mathcal M}\E[\rho(\tau)] \;\le\; \sup_{(F^\theta)\in\mathcal R}\ \int \rho\,\d\widetilde F. \label{eq:relaxation}
\end{align}

\begin{lemma}[Weak duality]\label{lem:duality}
Fix a nonnegative, decreasing, continuous $\rho$ and a path $\rp{\mu^*_t}$ with deterministic survivor belief $\widehat\mu_t$. Suppose there are $a\in\R^{|\Theta|}$ and a \emph{finite, atomless} nonnegative measure $\lambda$ on $\R_+$, with continuous tail $\Lambda(t):=\lambda((t,\infty))$ and $\Lambda(0)<\infty$, such that for every $\theta\in\Theta$ and every $t\in[0,\infty]$,
\begin{align}
\rho(t) + \chi\int_0^t \Lambda(s)\,\d s - \int_{(0,t)}\nabla H(\widehat\mu_s)\cdot e_\theta\,\d\lambda(s) - \Lambda(t)\,H(e_\theta) \;\le\; a\cdot e_\theta, \tag{Dual}\label{eq:dual}
\end{align}
the case $t=\infty$ read as the limit of the left-hand side (with $\Lambda(\infty)=0$). Then $\int\rho\,\d\widetilde F \le a\cdot\mu_0 + H(\mu_0)\,\Lambda(0)$ for every $(F^\theta)\in\mathcal R$, and the right-hand side does not depend on $(F^\theta)$. If $\rp{\mu^*_t}$ is feasible and resolves the state almost surely, \eqref{eq:cap} binds along $F^*$, and for each $\theta$ \eqref{eq:dual} holds with equality for $F^{*\theta}$-almost every $u$, then $\rp{\mu^*_t}$ attains the left supremum and $F^*$ the right supremum in \eqref{eq:relaxation}.
\end{lemma}
\begin{proof}{Proof.}
Let $(F^\theta)\in\mathcal R$ and let $B(t)\ge 0$ denote the slack in \eqref{eq:cap} at time $t$:
\begin{align*}
B(t) := \chi\!\int_0^t(1-F(s))\,\d s - \sum_i F^i(t)\,H(e_i) - H\Bigl(\sum_i\bigl(\mu_0(i)-F^i(t)\bigr)e_i\Bigr) + H(\mu_0).
\end{align*}
Since $\lambda\ge 0$, adding $\int B\,\d\lambda \ge 0$ can only increase the objective: $\int\rho\,\d\widetilde F \le \int\rho\,\d\widetilde F + \int_0^\infty B(t)\,\lambda(\d t)$.

The only term in $B$ that is nonlinear in $(F^\theta)$ is the entropy term. The homogeneous extension of $H$ is convex on the nonnegative orthant, so it dominates each of its supporting rays: for every $y$ in the orthant,
\begin{align*}
H(y) \;\ge\; \nabla H(\widehat\mu_t)\cdot y,
\end{align*}
since for $y\neq 0$, $H(y)=\lVert y\rVert_1\, H(y/\lVert y\rVert_1)\ \ge\ \lVert y\rVert_1\bigl[H(\widehat\mu_t)+\nabla H(\widehat\mu_t)\cdot(y/\lVert y\rVert_1-\widehat\mu_t)\bigr] = \nabla H(\widehat\mu_t)\cdot y$ by convexity on the simplex and Euler's identity, while at $y=0$ both sides vanish. Applying this at $y=\sum_i(\mu_0(i)-F^i(t))e_i$ and substituting $-\nabla H(\widehat\mu_t)\cdot y$ for the entropy term replaces $B$ by a larger quantity $B_{\mathrm{lin}}$ that is \emph{linear} in $(F^\theta)$.

Now integrate $B_{\mathrm{lin}}$ against $\lambda$ and regroup each term as an integral against $\d\widetilde F^\theta$ by Fubini--Tonelli. All exchanges are licensed: $\lambda$ is finite and atomless, $\nabla H(\widehat\mu_s)\cdot e_\theta$ and $H(e_\theta)$ are bounded along the path, and $1-F(s)=\sum_\theta\widetilde F^\theta((s,\infty])$:
\begin{itemize}
\item[--] $\displaystyle\int_0^\infty \chi\!\int_0^t (1-F(s))\,\d s\,\lambda(\d t) = \chi\!\int_0^\infty (1-F(s))\,\Lambda(s)\,\d s = \sum_\theta\int_{(0,\infty]}\Bigl[\chi\!\int_0^u\Lambda(s)\,\d s\Bigr]\d\widetilde F^\theta(u)$;
\item[--] $\displaystyle-\int_0^\infty \sum_\theta F^\theta(t)\,H(e_\theta)\,\lambda(\d t) = -\sum_\theta\int_{(0,\infty]} \Lambda(u)\,H(e_\theta)\,\d\widetilde F^\theta(u)$, using $F^\theta(t)=\widetilde F^\theta([0,t])$ for finite $t$ and $\Lambda(\infty)=0$ at the atom;
\item[--] $\displaystyle-\int_0^\infty \nabla H(\widehat\mu_t)\cdot\sum_\theta\bigl(\mu_0(\theta)-F^\theta(t)\bigr)e_\theta\,\lambda(\d t) = -\sum_\theta\int_{(0,\infty]}\Bigl[\int_{(0,u)}\nabla H(\widehat\mu_s)\cdot e_\theta\,\d\lambda(s)\Bigr]\d\widetilde F^\theta(u)$;
\item[--] $\displaystyle\int_0^\infty H(\mu_0)\,\lambda(\d t) = H(\mu_0)\,\Lambda(0)$, a constant independent of $(F^\theta)$.
\end{itemize}
Adding $\int\rho\,\d\widetilde F=\sum_\theta\int_{(0,\infty]}\rho(u)\,\d\widetilde F^\theta(u)$, the first three bullets assemble, against each $\d\widetilde F^\theta(u)$, exactly the left-hand side of \eqref{eq:dual} evaluated at $(\theta,u)$---at $u=\infty$, its limiting value. Hence
\begin{align*}
\int\rho\,\d\widetilde F \;&\le\; \sum_\theta\int_{(0,\infty]} \bigl[\text{LHS of \eqref{eq:dual}}\bigr]\,\d\widetilde F^\theta(u) + H(\mu_0)\,\Lambda(0)\\
&\le\; \sum_\theta a\cdot e_\theta\,\widetilde F^\theta\bigl((0,\infty]\bigr) + H(\mu_0)\,\Lambda(0) \;=\; a\cdot\mu_0 + H(\mu_0)\,\Lambda(0),
\end{align*}
where the second inequality applies \eqref{eq:dual} pointwise and the last equality uses $\widetilde F^\theta((0,\infty])=\mu_0(\theta)$.

Finally, when $\rp{\mu^*_t}$ resolves the state almost surely, $F^*$ places no mass at $u=\infty$, and at $F^*$ every inequality above binds: the slack $B$ vanishes $\lambda$-almost everywhere because \eqref{eq:cap} holds with equality along the greedy path; the supporting-ray inequality is tight because the survivor vector of $\rp{\mu^*_t}$ is proportional to $\widehat\mu_t$, the point at which $H$ was supported; and for each $\theta$, \eqref{eq:dual} holds with equality $F^{*\theta}$-almost everywhere by hypothesis. The upper bound $a\cdot\mu_0+H(\mu_0)\Lambda(0)$ does not depend on $(F^\theta)$, so $F^*$ attains the right supremum in \eqref{eq:relaxation}, and the feasible process $\rp{\mu^*_t}$, whose value equals $\int\rho\,\d\widetilde F^*$, attains the left.
\Halmos\end{proof}

\section{Proof of Proposition 1}\label{sec:proof-prop1}

I first show the greedy process is well-defined, then construct the dual certificate for strictly convex $C^{(2)}$ objectives (\cref{lem:attain}), and finally extend to all admissible $\rho$ by approximation (\cref{lem:moll}).

\paragraph{Well-definedness.} Within round $k$ with active set $\Theta^k$, \eqref{eq:isoD} reduces (using $\H H(\widehat\mu_t)\widehat\mu_t=\mathbf 0$) to $\Sigma(\widehat\mu_t)\,\bm\beta_t=\alpha\bm 1$, where $\Sigma(\mu)=[\H H(\mu)_{\theta\theta'}]_{\theta,\theta'\in\Theta^k}$. By \cref{ass:1}(iii) the full Hessian $\H H(\mu)$ has nonpositive off-diagonal entries, by convexity it is positive semidefinite, by homogeneity $\H H(\mu)\mu=\mathbf 0$ with $\mu\gg 0$, and $\H H(\mu)$ is irreducible by assumption; hence $\H H(\mu)$ is an \emph{irreducible singular symmetric $M$-matrix}. A standard property of such matrices is that every proper principal submatrix is a nonsingular $M$-matrix, hence positive definite \citep[Ch.~6]{berman-plemmons-1994}. Since $|\Theta^k|<n$, $\Sigma(\widehat\mu_t)$ is therefore positive definite with strictly positive inverse, and \eqref{eq:isoD}--\eqref{eq:info-rate} pin down a unique positive rate vector
\begin{align*}
\bm\beta_t = \frac{\chi\,\Sigma(\widehat\mu_t)^{-1}\bm 1}{[D(e_\theta\mid\widehat\mu_t)]^\top_{\theta\in\Theta^k}\cdot\Sigma(\widehat\mu_t)^{-1}\bm 1}.
\end{align*}
For Lipschitz continuity I confine the dynamics to a compact subset of the open simplex. Assume without loss that $\mu_0$ has full support (null states can be dropped). Along the greedy path the active sets grow, $\Theta^k\subseteq\Theta^{k+1}$, and the two kinds of coordinates are controlled separately. An \emph{inactive} coordinate $i\notin\Theta^k$ is nondecreasing within the round, since \eqref{eq:drift} gives $\d\widehat\mu_t(i)/\d t=\widehat\mu_t(i)\sum_\theta\beta_t(\theta)\ge 0$; it therefore stays at or above $\mu_0(i)$ until it activates. An \emph{active} coordinate stays above $\epsilon$: a member of the closest set falling to $\epsilon$ would contradict \cref{ass:1}(ii). Every coordinate thus remains at or above $\delta_0:=\min\{\epsilon,\min_i\mu_0(i)\}>0$, so the trajectory lies in the compact set $K_{\delta_0}:=\{\mu\in\Delta(\Theta):\mu(i)\ge\delta_0\ \forall i\}\subset\Delta(\Theta)^\circ$, which is bounded away from every vertex. On $K_{\delta_0}$ the entries of $\Sigma$, of $\Sigma^{-1}$, and the denominator above are bounded with the denominator bounded away from zero, and $\zeta(t):=\min_\theta D(e_\theta\mid\widehat\mu_t)$ is bounded away from zero ($D(e_\theta\mid\cdot)$ is continuous and vanishes only at $e_\theta$); so $\widehat\mu_t\mapsto\bm\beta_t$ is Lipschitz, and with \eqref{eq:drift}, Picard--Lindel\"of gives a unique $C^{(1)}$ path. A transition occurs in finite time: $\max_\theta\beta_t(\theta)\ge\delta:=\chi/\sup_\mu\min_\theta D(e_\theta\mid\mu)>0$ by \cref{ass:1}(i), so absent a transition some active coordinate would fall to $\epsilon$ in bounded time, contradicting \cref{ass:1}(ii); finiteness of $\Theta$ ends the recursion at some round $K\le n$.

\begin{lemma}[Optimality for strictly convex objectives]\label{lem:attain}
Let $\rho:\R_+\to\R_+$ be positive, decreasing, strictly convex and $C^{(2)}$. Then $\rp{\mu^*_t}$ attains $\sup_{\rp{\mu_t}\in\mathcal M}\E[\rho(\tau(\mu))]$.
\end{lemma}
\begin{proof}{Proof.}
I construct a certificate $(a,\lambda)$ satisfying the hypotheses of \cref{lem:duality} along the greedy path. Let $\zeta(t):=\min_\theta D(e_\theta\mid\widehat\mu_t)=D(e_\theta\mid\widehat\mu_t)$ for $\theta\in\Theta^k$ (Lipschitz, and bounded away from zero on $K_{\delta_0}$). Define $\Lambda$ by
\begin{align*}
\Lambda'(t)=\bigl(\rho'(t)+\chi\Lambda(t)\bigr)/\zeta(t),\qquad \Lambda(\infty)=0,\qquad\text{i.e.}\quad \Lambda(t)=\int_t^\infty e^{\int_s^t \chi/\zeta(z)\,\d z}\,\frac{-\rho'(s)}{\zeta(s)}\,\d s,
\end{align*}
and set $\lambda:=-\Lambda'$. Since $\zeta>0$, $\lambda$ has the sign of $-\rho'(t)-\chi\Lambda(t)$, and
\begin{align*}
\chi\Lambda(t)=\int_t^\infty \chi\,e^{\int_s^t\chi/\zeta\,\d z}\,\frac{-\rho'(s)}{\zeta(s)}\,\d s < -\rho'(t)\int_t^\infty \chi\,\frac{e^{\int_s^t\chi/\zeta\,\d z}}{\zeta(s)}\,\d s = -\rho'(t),
\end{align*}
using strict convexity of $\rho$ ($-\rho'$ strictly decreasing, so $-\rho'(s)<-\rho'(t)$ for $s>t$) and $\int_t^\infty\frac{\d}{\d s}[-e^{\int_s^t\chi/\zeta\,\d z}]\,\d s=1$. Hence $\lambda>0$. Differentiating the left-hand side of \eqref{eq:dual} in $t$,
\begin{align*}
\frac{\d}{\d t}\Bigl[\rho(t)+\chi\!\int_0^t\!\Lambda - \Lambda(t)H(e_\theta) - \!\int_{(0,t)}\!\nabla H(\widehat\mu_s)\!\cdot e_\theta\,\d\lambda\Bigr] = \rho'(t)+\chi\Lambda(t)+\lambda(t)\,D(e_\theta\mid\widehat\mu_t),
\end{align*}
which is zero for $\theta\in\Theta^k$ at $t\ge\widehat t^k$ (by the definition of $\Lambda$, since $D(e_\theta\mid\widehat\mu_t)=\zeta(t)$ there) and strictly positive otherwise. Setting each $a\cdot e_\theta$ to the common value of the left-hand side at $t=\widehat t^K$, \eqref{eq:dual} therefore holds for every $t\in[0,\infty]$: for $\theta$ active at $t$ the left-hand side is constant and equal to $a\cdot e_\theta$; for $\theta$ not yet active it lies strictly below while nondecreasing; and its $t\to\infty$ limit equals $a\cdot e_\theta$ for every $\theta$, all states being active from $\widehat t^K$ on. In particular, for each $\theta$, \eqref{eq:dual} holds with equality on $\mathrm{supp}\,F^{*\theta}\subseteq[\widehat t^{k(\theta)},\infty)$, where $\widehat t^{k(\theta)}$ is $\theta$'s activation time. The multiplier qualifies for \cref{lem:duality}: $\zeta$ is Lipschitz and bounded away from zero along the path, so $\Lambda$ is $C^{(1)}$ with $\Lambda(0)<\infty$, and $\lambda=-\Lambda'$ is a finite atomless measure. Constraint \eqref{eq:cap} binds along the greedy path because \eqref{eq:info-rate} makes the throughput constraint hold with equality at every $t$, and the greedy process resolves the state almost surely, its terminal-phase breakthrough hazard $\chi/\zeta$ being constant and positive. The hypotheses of \cref{lem:duality} are met, so $\rp{\mu^*_t}$ attains the relaxed---hence, by \eqref{eq:relaxation}, the original---supremum.
\Halmos\end{proof}

\begin{lemma}[Extension by mollification]\label{lem:moll}
\cref{lem:attain} extends to every nonnegative, weakly decreasing, continuous, weakly convex $\rho:\R_+\to\R_+$.
\end{lemma}
\begin{proof}{Proof.}
Fix $\eta>0$. Convolve $\rho$ (extended to $\R$ along its left tangent at $0$, which preserves convexity and monotonicity) with a symmetric mollifier supported on $[-h,h]$: the result $\rho_h$ is $C^\infty$, convex (Jensen), weakly decreasing ($\rho_h'=\rho'*\varphi\le 0$), and nonnegative ($\rho_h\ge\rho\ge 0$ on $\R_+$ by symmetry and convexity), and $\rho\le\rho_h\le\rho+\tfrac\eta2$ for $h$ small (uniform continuity). Add $\tfrac\eta2 e^{-t}$---positive, decreasing, strictly convex---to obtain $\rho_\eta:=\rho_h+\tfrac\eta2 e^{-t}\in[\rho,\rho+\eta]$, now strictly convex, $C^{(2)}$, positive, and decreasing. By \cref{lem:attain}, $\E[\rho_\eta(\tau(\mu^*))]=\sup_\mu\E[\rho_\eta(\tau(\mu))]$, so
\begin{align*}
\sup_\mu\E[\rho(\tau(\mu))]\le\sup_\mu\E[\rho_\eta(\tau(\mu))]=\E[\rho_\eta(\tau(\mu^*))]\le\E[\rho(\tau(\mu^*))]+\eta.
\end{align*}
Letting $\eta\to0$ proves the claim, which is \cref{prop:1}.
\Halmos\end{proof}

\section{Assumption 1 for the Canonical Generators}\label{sec:appendix-assumption}

\paragraph{Kullback--Leibler.} For Shannon entropy the degree-one homogeneous extension is $H(x)=\sum_i x_i\log\bigl(x_i/\sum_j x_j\bigr)$, with $\H H(\mu)=\mathrm{diag}(1/\mu)-\bm 1\bm 1^\top$ on the simplex; thus $\H H(\mu)\mu=\bm 1-\bm 1=\mathbf 0$, the off-diagonal entries equal $-1$ (so $\H H$ is irreducible, as the maintained assumption requires), and for $\theta\ne\theta'$,
\begin{align*}
(e_\theta-\mu)^\top\nabla_\mu D(e_{\theta'}\mid\mu)= -(e_\theta-\mu)^\top\H H(\mu)(e_{\theta'}-\mu)
= -(e_\theta-\mu)^\top\bigl[\mathrm{diag}(1/\mu)-\bm 1\bm 1^\top\bigr](e_{\theta'}-\mu).
\end{align*}
Since $\bm 1^\top(e_{\theta'}-\mu)=0$, the rank-one term drops, and a direct computation gives $(e_\theta-\mu)^\top\mathrm{diag}(1/\mu)(e_{\theta'}-\mu)= -(1-\mu_\theta)-(1-\mu_{\theta'})+\sum_{i\neq\theta,\theta'}\mu_i = -1$; with the leading minus sign, the \cref{ass:1}(iii) value $(e_\theta-\mu)^\top\nabla_\mu D(e_{\theta'}\mid\mu)=\mathbf{+1}$ for all interior $\mu$ and all $\theta\ne\theta'$. Thus \cref{ass:1}(iii) holds with margin in every dimension; \cref{ass:1}(i)--(ii) hold because $D(e_\theta\mid\mu)=-\log\mu_\theta$.

\paragraph{Quadratic (Mahalanobis), binary only.} For $H(\mu)=\tfrac12\lVert\mu\rVert^2$ (homogenized), \cref{ass:1}(iii) reads $(e_\theta-\mu)^\top(e_{\theta'}-\mu)\le 0$. When $n=2$ the two vertex directions are anti-parallel---$e_2-\mu=-(\mu_1/\mu_2)(e_1-\mu)$---so $(e_1-\mu)^\top(e_2-\mu)=-(\mu_1/\mu_2)\lVert e_1-\mu\rVert^2\le 0$ holds automatically: \cref{ass:1}(iii) \emph{has no bite at $n=2$}, for any convex $H$. For $n\ge 3$ it can fail. Hence the quadratic generator enters only through the binary examples, where $n=2$.

\paragraph{Monotone stopping density.} For both generators the greedy minimal divergence $\zeta(t)=\min_\theta D(e_\theta\mid\mu^*_t)$ is non-decreasing along the path: on a single-active-state segment $\d\zeta/\d t = \beta_\theta(e_\theta-\mu)^\top\H H(\mu)(e_\theta-\mu)\ge0$ by positive semidefiniteness, so the breakthrough hazard $\chi/\zeta$ is non-increasing and the stopping density $f^*$ weakly decreasing---condition~(a) of \cref{sec:optmech}.

\end{APPENDICES}
\newpage
\setlength{\bibsep}{-1pt}
\bibliography{references}

\end{document}